# Characterisation of Corrosion Damage in T91/F91 steel exposed to Liquid Lead-Bismuth Eutectic

M.T. Lapington[1, *], M. Zhang[2], M.P. Moody[2], W.Y. Zhou[3], M.P. Short[3], F. Hofmann[1,^]

[1]Department of Engineering Science, University of Oxford, Parks Road, Oxford OX1 3PJ, UK

[2]Department of Materials, University of Oxford, Parks Road, Oxford OX1 3PH, UK

[3]Department of Nuclear Science & Engineering, Massachusetts Institute of Technology, 77 Massachusetts Avenue, Cambridge, MA 02139, USA

ABSTRACT

T91 samples were exposed to static liquid lead-bismuth eutectic (LBE) at 700°C for 250-500 hours in either an oxidising or reducing environment. Corrosion damage was characterised using electron microscopy techniques, which identified networks of LBE intrusion beneath LBE-wetted surfaces. Under reducing conditions these networks are uniformly distributed, while they appear patchier and deeper under oxidising conditions. The individual intrusions preferentially follow microstructural features, initially along prior-austenite grain boundaries, followed by penetration down martensite lath boundaries. Local depletion of Cr was observed within 4 µm of the intrusions and along intersecting boundaries, suggesting local Cr dissolution as the main corrosion mechanism.

GRAPHICAL ABSTRACT

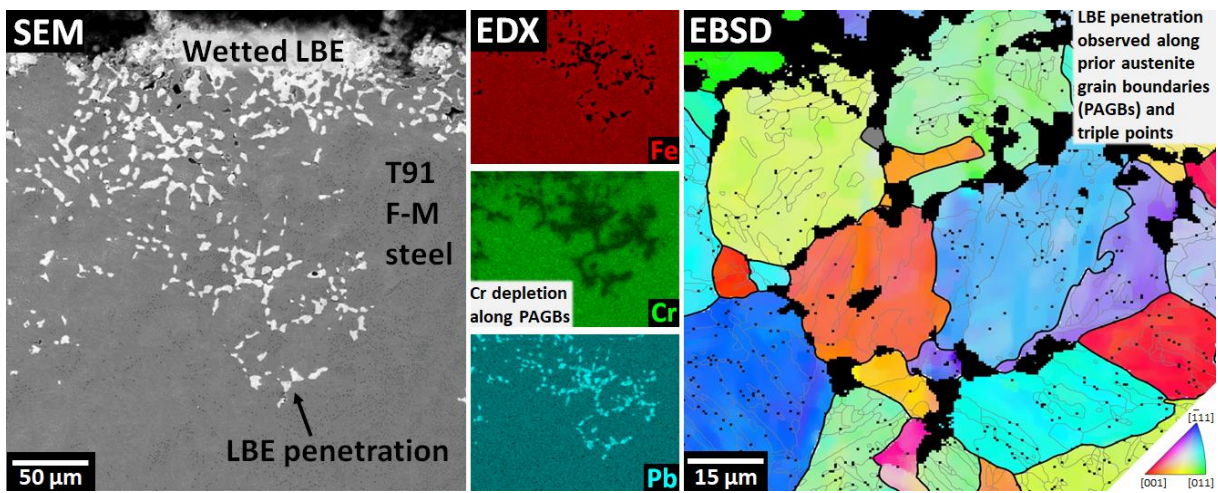

[*]Corresponding Author; email: mark.lapington@eng.ox.ac.uk

[^]email: felix.hofmann@eng.ox.ac.uk

______________________________________________________________________________

Nuclear fission remains a vital source of low carbon footprint energy, with much of the current fission reactor infrastructure in desperate need of replacement [1]. Liquid lead-cooled Generation-IV fast reactor designs promise exceptional power density and economic benefits without the inherent risks of reactive liquid sodium or radioactive molten salts. [2]. Maximising the efficiency of these reactor designs incentivises increasing reactor operating temperatures, which are currently limited to around 550°C due to corrosion-related concerns.

To reach a more economically viable target operating temperature of 700°C, any liquid metal facing materials must maintain a high tensile strength whilst resisting the highly corrosive liquid lead-bismuth eutectic [3–5]. For these reasons, materials that can generate and maintain a passivating oxide layer at elevated temperatures, combined with dissolution of oxygen into the melt is highly desirable [6]. The current 550°C design is limited by the iron-based oxide layer that transforms into non-protective Wüstite at 570°C [7]. Any high temperature material candidate must be able to form dense protective oxides even in areas of low oxygen potential, to prevent the formation of potentially catastrophic pits and crevices. Some alternative oxide-forming elements include Cr, Si, and Al, all of which form dense protective oxides at lower oxygen potentials than iron [5,6,8,9]. Much research into Cr, Si and Al-containing steels has been carried out, but so far none of these candidates have managed to combine all of the requirements of workability, structural strength and corrosion resistance [5,10].

One promising alternative is to apply a corrosion-resistant Fe-Cr-Si steel as an LBE-facing barrier, bonded to a sturdier structural steel in the form of a functionally-graded composite [10]. Short et al. selected the ferritic/martensitic alloy T91 (also referred to as F91 or Grade 91) as a corrosion and radiation-resistant structural steel, due to its history of use in the fossil fuel sector [11–13]. Further experiments performed on model Fe-Cr alloys exposed to molten salts [14] showed that corrosion damage differed from grain to grain, suggesting orientation-dependent behaviour. This study aims to investigate the response of T91 to immersion in liquid LBE, focusing on the orientation dependence of corrosion to better predict its extent and establish its underlying mechanism.

Billets of T91 were purchased in the quenched and tempered condition, then heat treated as per manufacturer's specification [12,15]. The bulk composition (in wt.%) has been verified by inductively-coupled plasma mass spectroscopy (ICP-MS) [15] and atom probe tomography (APT) [16], as shown in Table 1.

|        | Fe   | Cr   | Si   | W     | Mo   | Cu   | Ni   | Mn   | V    | C     | N     | P     | S     | Al    | Nb    |
|--------|------|------|------|-------|------|------|------|------|------|-------|-------|-------|-------|-------|-------|
| ICP-MS | 89.0 | 8.39 | 0.32 | 0.008 | 0.92 | 0.15 | 0.28 | 0.47 | 0.21 | 0.105 | 0.057 | 0.019 | 0.003 | 0.006 | -     |
| APT    | 89.7 | 7.91 | 0.39 | -     | 0.88 | 0.11 | 0.30 | 0.48 | 0.11 | 0.002 | 0.013 | 0.009 | -     | 0.001 | 0.001 |

*Table 1 - The bulk composition of as-received T91 used in this study (in wt.%) as verified by ICP-MS [15], and an average composition obtained from six atom probe datasets [16].*

Samples of T91 were prepared and placed inside alumina crucibles filled with molten LBE, heated by upright cylindrical furnaces under a shroud of cover gas inside a sealed autoclave (Diagrams available in supplementary material). Each chamber was fed with a separate cover gas to produce an oxidising or reducing environment with respect to the potential for the formation of iron oxide. Oxygen levels for the oxidising and reducing conditions were maintained at around $1\times10^{-20}$ atm and below $1\times10^{-24}$ atm respectively corresponding to oxide formation potentials above and below FeO at the testing temperature. The exposure consisted of lowering the sample into the LBE crucible for exposure times of 245 h or 506 h at temperatures of 700°C or 715°C respectively.

To focus on grain orientation, the LBE-exposed samples were polished at a 5° gradient from the exposed surface. This involved mounting a sample onto a polishing holder using a shim to ensure a 5° angle, as shown in Fig. 1a with a schematic shown in Fig. 1b. Lacomit varnish was used to protect the sample surface during grinding and polishing with diamond paste and colloidal silica. The sample was remounted onto a microscopy stub using conductive epoxy to reduce charging effects. This sample preparation method preserves some of the samples original LBE-wetted surface (see Figs. 1c and d), and the position of the LBE/metal interface which allows estimates of LBE penetration depth using the 1:11 Y-value-to-depth ratio calculated in Fig. 1b. This technique was further refined to allow easier handling and more consistent polishing, by hot mounting the sample subassembly inside a phenolic resin puck (see Fig. 1e).

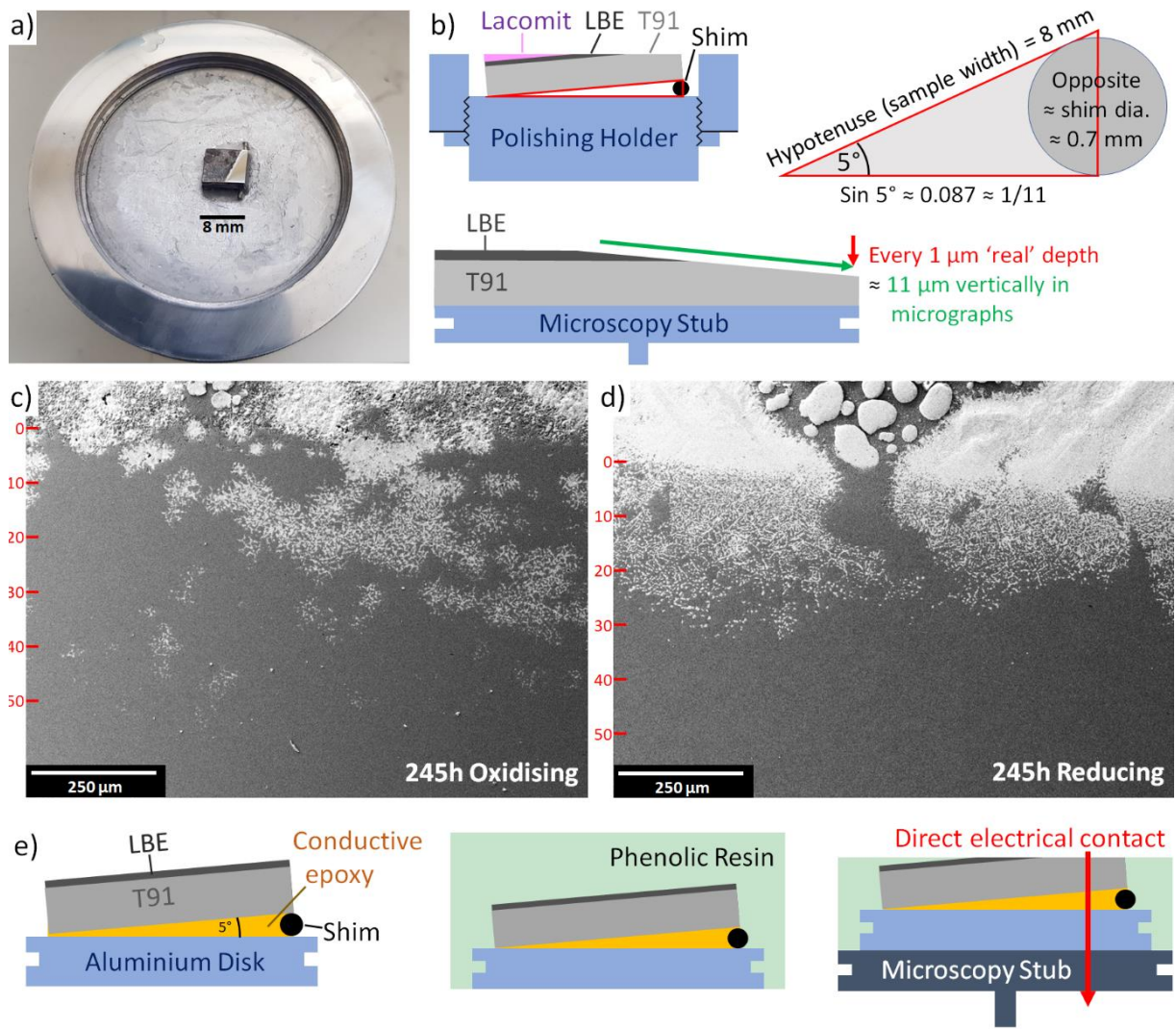

*Figure 1 – Gradient polishing technique: (a) Polishing holder with mounted and part-polished sample. (b) Schematic of sample mounting for initial gradient polishing. SEM micrographs: (c) 245 h oxidising atmosphere (245h-OX) with red depth markers. (d) 245 h Reducing atmosphere (245h-RED) with red depth markers. (e) Revised gradient polishing technique, illustrating the sample mounting (left), encasing in phenolic resin (middle) and mounting of the final, gradient-polished sample (right). Note that (b) and (e) are not to scale.*

Polished samples were analysed either in a Zeiss Merlin scanning electron microscope (SEM) with a Bruker Quantax electron backscatter diffraction (EBSD) detector, or a Zeiss Crossbeam 540 with an Oxford Instruments XmaxN150 energy-dispersive X-ray spectroscopy (EDX) detector and a Nordlys Max EBSD detector. The sample exposed to an oxidising atmosphere for 245 h (245h-OX) has a rough surface texture indicative of the presence of an oxide layer (Fig. 1c). The LBE coverage is patchy, likely due to differing wetting characteristics between oxide and metal. A small non-wetted area with LBE droplets on the surface can be seen left of the centre. In contrast, the areas where LBE wets the surface correlate with the presence of subsurface LBE infiltration, which resemble the martensite lath networks in scale and morphology. On the sample exposed to a reducing atmosphere for 245 h (245h-RED) most of the surface is wetted, presumably due to the relative absence of a protective oxide layer (Fig. 1d). The distribution of LBE infiltration is more homogeneous, although shallower compared to 245h-OX. Areas of dewetted LBE are also present, with comparitively larger LBE droplets.

Additional EDX micrographs were obtained to image elemental distributions, using a 1 nA 20 kV electron beam to avoid overlap between the Cr and O spectral peaks. Fig. 2 shows SEM and EDX maps of 245h-OX and 245h-RED at magnifications of 100x and 500x. Separate EDX maps are shown for Fe (red), Pb (cyan), and Cr (green), with a larger composite plot showing all three elements. A sharp interface exists between LBE intrusions and Fe matrix, as seen in the higher magnification micrographs (Figs. 2b and d). The Cr maps and combined EDX maps also reveal a 3-4 µm wide Cr-depletion zone adjacent to the LBE intrusions, indicating leaching of Cr from the matrix. It is impossible to tell whether this is due to the formation of Cr-rich phases elsewhere (eg: carbides [13] or oxides [17,18]), or just dissolution into the LBE. Other Cr-rich phases were noticeably absent in these EDX maps.

The pattern of LBE infiltration networks matches the SEM observations in Fig. 1, with noticeable differences between the two atmospheres: Under oxidising conditions (245h-OX), patchy 'snowflake' patterns form that penetrate up to 50 µm beneath the surface (Figs. 2a and b). Under reducing conditions (245h-RED), the network is more uniform and extends to a shallower depth of 30 µm beneath the surface. Neither condition shows appreciable shrinking or thinning of the intrusion fingers, as would be typical of grain boundary oxidation [19].

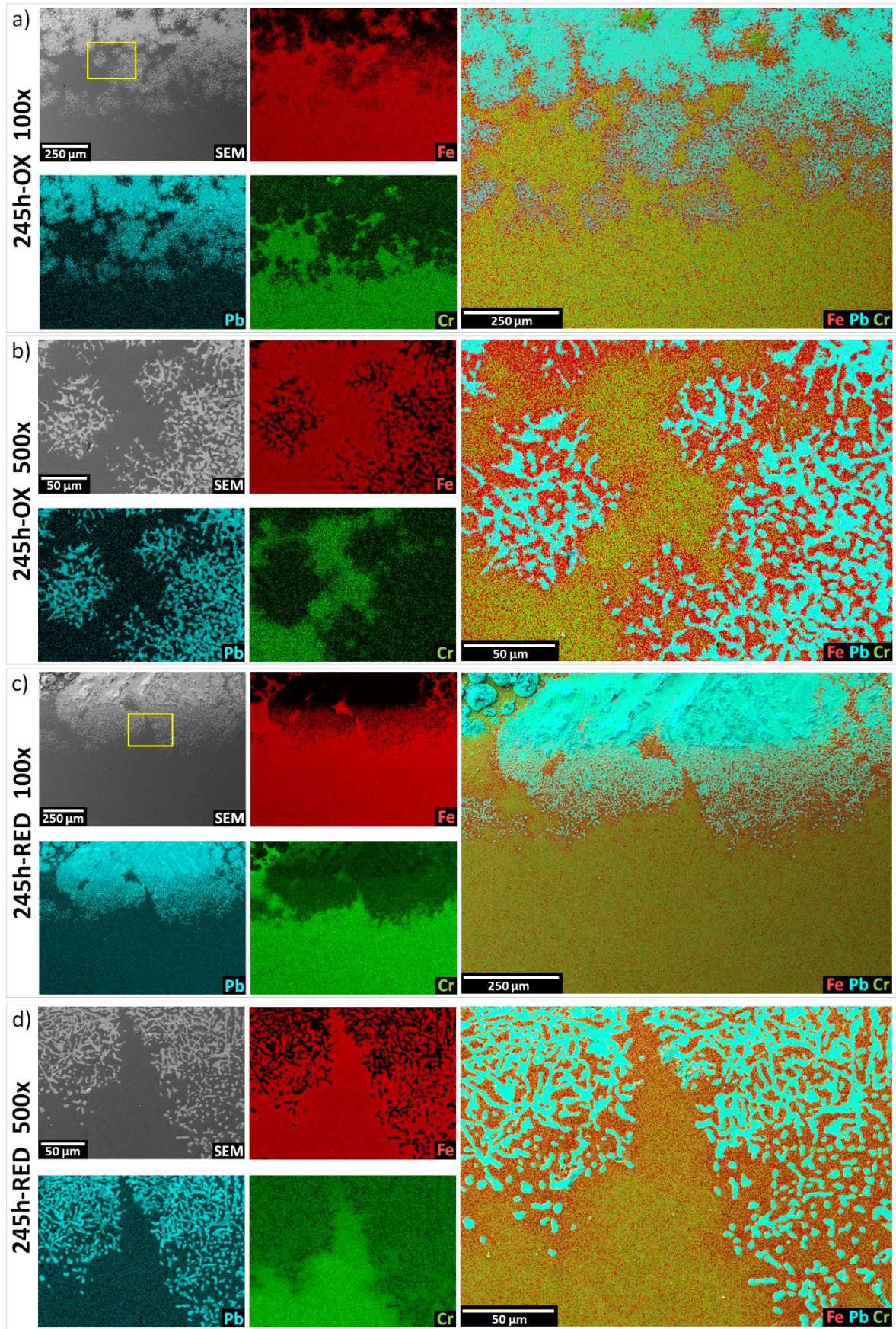

Figure 2 – SEM/EDX maps showing elemental distributions of Fe (red), Pb (cyan) and Cr (green): (a) 245h-OX 100x mag, (b) 245h-OX 500x mag, (c) 245h-RED 100x mag, (d) 245h-RED 500x mag.

EBSD micrographs were obtained to investigate the relationship between corroded areas and grain orientation, boundary character, or the presence of other phases. EBSD data were processed using the MTEX software suite [20] to analyse the relative misorientation of all grain boundaries. Prior-austenite grains (PAGs) undergoing a martensitic transition share an orientation relationship with the resulting lath microstructure [21,22], which allows the network of prior austenite grain boundaries (PAGBs) to be recovered and analysed in MTEX using the techniques reported by Niessen et al. [23].

Fig. 3 shows correlative micrographs from samples 245h-OX and 245h-RED. The SEM images on the left and corresponding EDX maps below show the composition of the samples, while the EBSD maps on the right are coloured according to the average crystallographic orientation of each reconstructed PAG. Black non-indexed regions represent the LBE. Grey areas represent 'orphan' grains that MTEX could not match to a PAG, tending to be round, lath-free, and adjacent to LBE intrusions. MTEX calculated grain boundaries are overlaid, with PAGBs shown as thick black lines and lath boundaries shown as thinner grey lines.

The maps for 245h-OX (Fig. 3a) reveals LBE infiltration along grain boundaries, with depletion of Cr shown in EDX. The MTEX-processed EBSD scan on the right predicts PAGBs that coincide with LBE intrusions and Cr depletion in the EDX map. Observations of LBE infiltrations at PAGBs and triple junctions suggests that infiltration is aided by disordered boundaries. Reducing conditions (Fig. 3b) exhibit different behaviour; EDX maps still show Cr depletion around the LBE intrusions, but no Cr depletion is evident along grain boundaries. Additionally, the correlation between reconstructed PAGBs and LBE intrusions is less clear, especially for the smaller LBE network seen in the top left of the EBSD map which is centred on the grain boundary but extends well into the grains on either side.

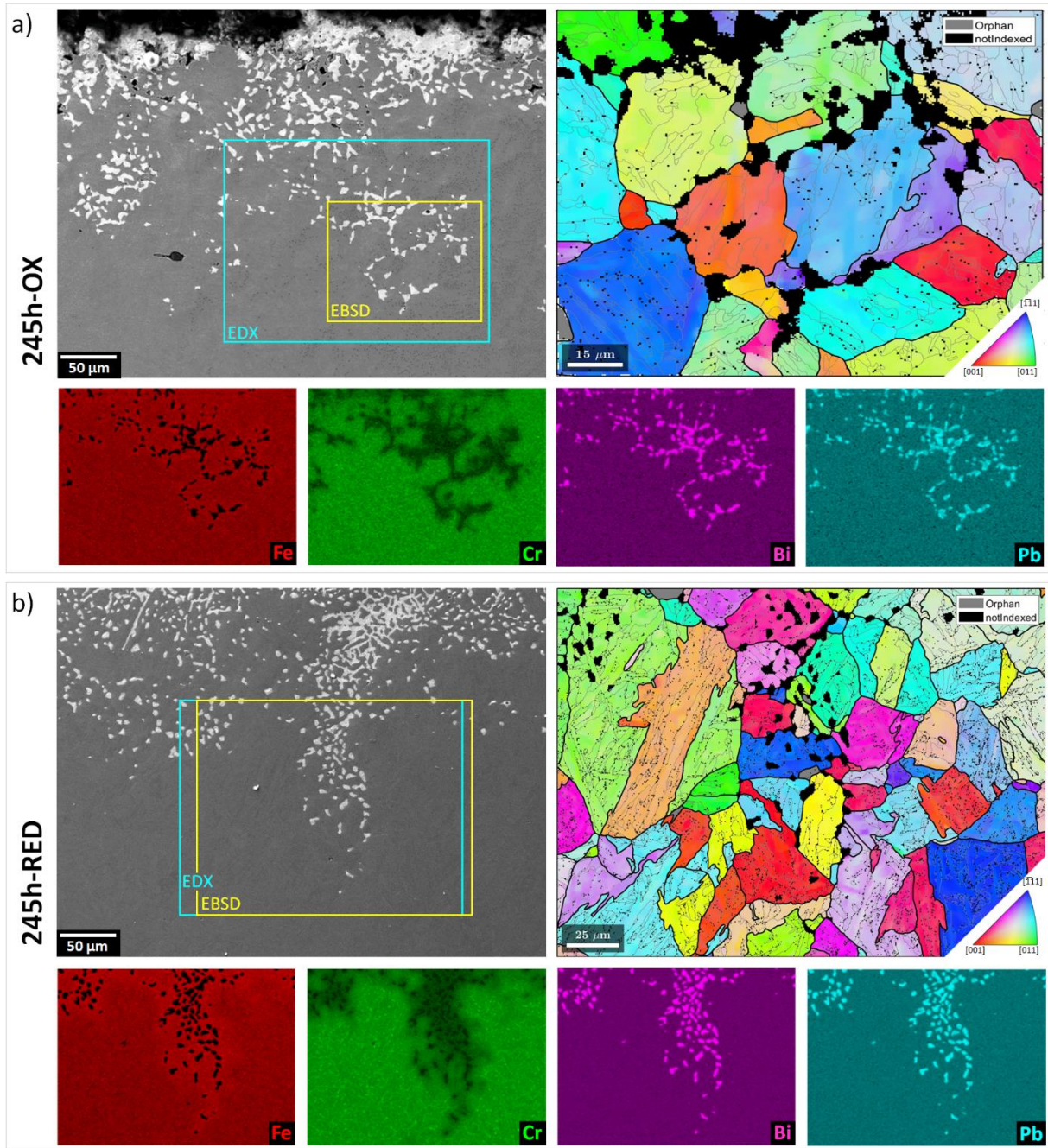

*Figure 3 – 245 hour exposure SEM (left), EDX (below), and EBSD maps post-processed in MTEX to show average orientations of parent austenite grains coloured by IPF (right): (a) Oxidising (b) Reducing. EDX and EBSD maps taken from marked boxes. Black non-indexed regions represent LBE, while grey regions represent 'orphan' grains that could not be assigned to a parent austenite grain. Grain boundaries plotted; black = PAGB, grey = laths.*

Micrographs from the samples exposed to LBE for 506 h in the oxidising (506h-OX) and reducing (506h-RED) environments are shown in Figs. 4a and b respectively. The SEM micrographs show that LBE intrusions have progressed much further into the material (approximately 60 μm for 506h-OX and 40 µm for 506h-RED). The LBE intrusion networks follow the same general trends as the 245 h exposures (Fig. 3), with the oxidising environment leading to deeper networks with inconsistent distribution, and the reducing environment producing a more uniform but shallower network. Significant corrosion can be seen at PAGBs and triple points. Focussing on the EDX maps in Fig. 4a, the same pattern of Cr depletion can be seen in areas adjacent to LBE intrusions, especially along MTEX-calculated PAGBs. In addition, Fig. 4a contains the largest number of round, lath-free orphan grains seen in any of the samples.

Safe application of T91-based materials for use in LBE-cooled reactors requires a thorough understanding of their corrosion characteristics at all points within the cooling loop [6]. In particular, the reliance on dissolved oxygen within the LBE itself to form a protective oxide layer depends on understanding how corrosion progresses, especially when the partial pressure of oxygen can vary due to temperature changes and flow characteristics [24,25]. The SEM micrographs shown in Figs. 1-4 reveal that the morphology of LBE intrusion networks is dependent on the oxygen content of the LBE. Exposure to a reducing atmosphere correlates with consistent LBE intrusion networks with a uniform depth, while an oxidising atmosphere leads to inconsistent patchy LBE networks. It is important to note that the deeper networks formed in the oxidising condition combined with the inconsistent patchiness leads to the hypothesis that the atmosphere dictates the location more than the volume of affected regions.

The observation that LBE intrusion networks tend to occur underneath areas of LBE wetted onto the surface is demonstrated well in Figs. 1c and d which cover both exposure conditions. This is consistent with previous observations of T91 exposed to LBE by Short et al. for 500 h exposures [15], and Gnecco at al. for 2000 h exposures [26]. Fig. 1c shows the surface roughness of the oxidising exposure, while the reducing exposure surface in Fig. 1d is uniformly smooth, even in areas of LBE wetting. Giuranno et al. have discussed the difficulty that LBE has wetting onto surface oxides [27]. The surface EDX maps (Figs. 2a and c) show enrichment of Cr and O, which indicates the presence of surface oxides in non-wetted regions. The oxygen maps have only been included as supplementary material, as it is unknown whether these oxides were present during the LBE exposure experiments or formed afterwards.

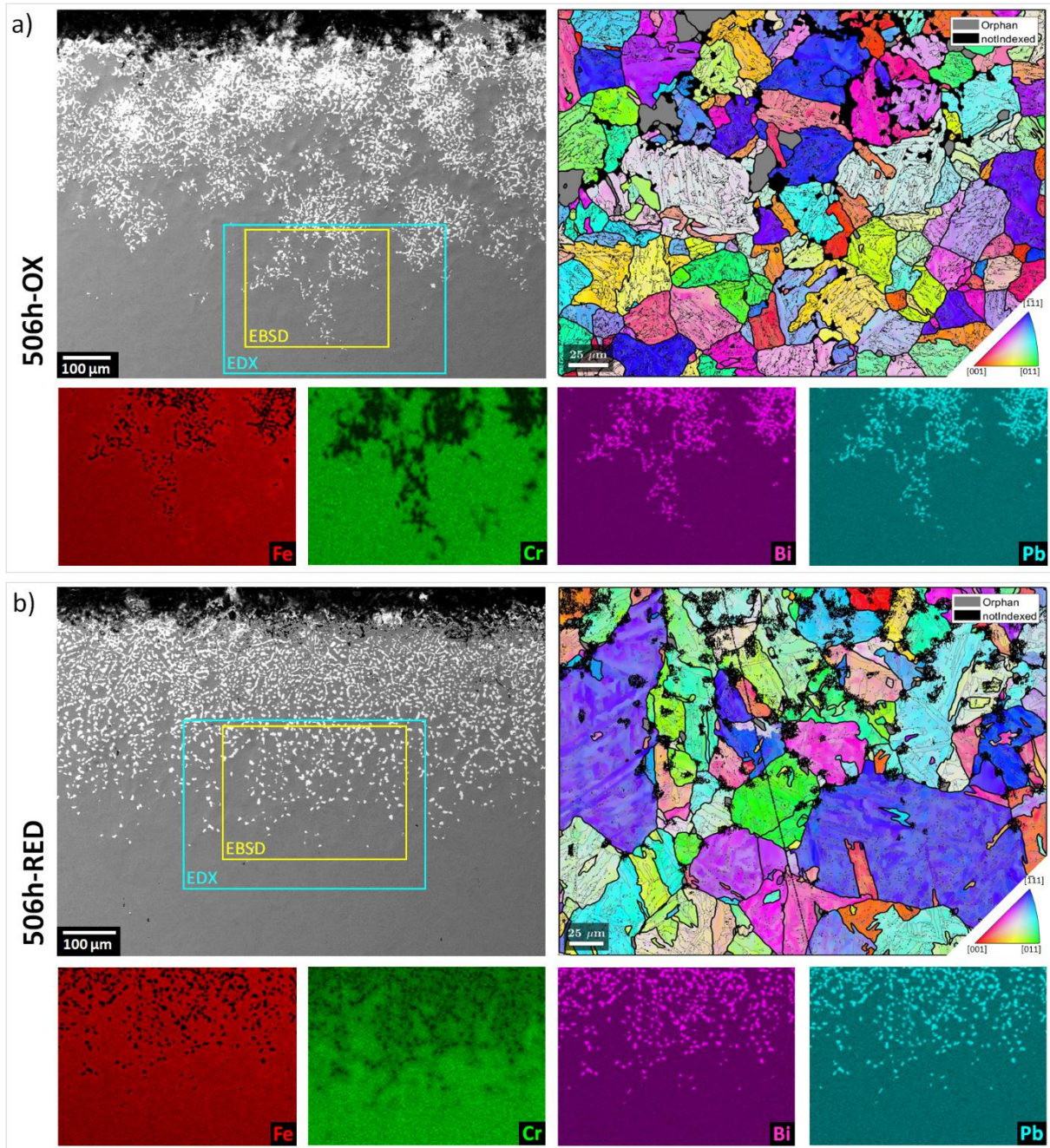

*Figure 4 – 506 hour exposure SEM (left), EDX (below), and EBSD maps post-processed in MTEX to show average orientations of parent austenite grains coloured by IPF (right): (a) Oxidising (b) Reducing. EDX and EBSD maps taken from marked boxes. Black non-indexed regions represent LBE, while grey regions represent 'orphan' grains that could not be assigned to a parent austenite grain. Grain boundaries plotted; black = PAGB, grey = laths.*

Considering individual LBE intrusion networks in more detail, the EDX maps (Figs. 2b and d) reveal clear Cr depletion zones extending up to 4 µm, similar to that seen during sensitisation of stainless steels due to carbide formation [28]. However the observed Cr depletion distance is an order of magnitude larger [29], likely due to continuous leaching. The driving force behind this localised Cr depletion is likely a combination of the high chemical potential for dissolution in LBE [4,30], combined with the faster diffusion rate of Cr relative to Fe through BCC Fe-Cr systems as shown in *ab initio* models [31,32].

The EDX maps (Figs. 3 and 4) show this Cr depletion at a higher magnification, with both oxidising samples exhibiting significant Cr leaching from intrusion adjacent boundaries. The reconstructed PAGBs closely coincide with the Cr-depleted boundaries, consistent with results by Haruo et al. on heat affected zones in low-carbon martensitic steels [33]. Most importantly, the network morphology changes between the dense surface networks and the single intrusions deeper in towards the bottom of the micrographs. It appears that LBE intrusions prefer to follow PAGBs and triple junctions at first, losing this preference during the later stages of corrosion, in agreement with observations by Short et al. and Tsisar et al. [15,25].

Several different mechanisms could be responsible for the preferential dissolution of PAGBs and triple junctions. They both exhibit an above average number of defects that would quicken dissolution [34]. There are also numerous observations of $M_{23}C_6$ carbides (where M = Fe or Cr) residing at PAGBs in ferritic-martensitic steels [25,35–37], which have been observed by Liu et al. to dissolve due to Cr removal, though this was attributed to the formation of Cr-rich oxides rather than direct Cr dissolution [38]. It is probable that a combination leads to the preferential LBE attack of these regions.

The presence of a round, lath-free 'orphan' grains in the EBSD datasets in Figs. 3 and 4 is also of interest. Their origin is unknown, some possiblities including; allotriomorphic ferrite found on PAGBs [39,40], transformed retained austenite, recrystallised ferrite due to dealloying similar to observations by Sapundjiev et al. [17], or possibly dissolution of lath boundary-pinning carbides. Their location adjacent to LBE intrusions implicates Cr leaching, dealloying, or dissolution of lath boundary carbides for their formation.

When considering individual LBE intrusions, the uniform thickness of the intrusions and lack of thinning, combined with the Cr depleted zones surrounding LBE intrusions shown in Fig. 2, suggests localised Cr dissolution as the predominant corrosion mechanism. We hypothesise that preferential Cr dissolution from PAGBs or triple points creates corrosion channels, which are backfilled by LBE to produce the infiltration networks. The characteristic width of the intrusions is likely a function of Cr diffusion distance in T91, and therefore be related to the observed 3-4 µm wide Cr depletion zone seen in Fig. 2. As a rough approximation, full Cr dissolution from this depletion zone should result in a 700 nm intrusion width, indicating that dissolution of the Fe matrix must also occur.

- T91 samples were exposed to LBE in oxidising or reducing atmospheres for 245-506 hours. The distribution of surface-wetted LBE differed, likely due to the formation of partially protective oxides under the oxidising atmosphere.
- LBE formed corrosion networks underneath surface-wetted LBE, leading to distinctive differences in distribution of LBE networks. The oxidising atmosphere caused less consistent and deeper LBE intrusions, while a shallower more uniform intrusion network was associated with reducing conditions.
- Cr dissolution was observed in EDX micrographs within ~4 µm of LBE intrusion networks and intersecting grain boundaries, suggesting local Cr dissolution as the main corrosion mechanism.
- Preferential attack of grain boundaries and triple points was observed in EBSD maps, which were identified as prior austenite grain boundaries after post-processing in MTEX.
- This relationship between preferential corrosion paths and microstructure is particularly important since grain boundary morphology can be modified by material processing, potentially opening up routes for improving material corrosion resistance.

ACKNOWLEDGMENTS

The authors acknowledge funding from the U.K. Engineering and Physical Sciences Research Council (EPSRC) through grant EP/T002808/1. Sample exposures at MIT were funded through the Nuclear Energy University Program (NEUP) grant 19-16754.

The authors acknowledge use of characterisation facilities within the David Cockayne Centre for Electron Microscopy, Department of Materials, University of Oxford, alongside financial support provided by the Henry Royce Institute (Grant ref EP/R010145/1). The Zeiss Crossbeam FIB/SEM used in this work was supported by EPSRC through the Strategic Equipment Funding grant EP/N010868/1. All raw data and scripts used in the production of this paper have been made available as a repository on Zenodo